\documentclass[runningheads]{llncs}
\usepackage[T1]{fontenc}
\usepackage{graphicx}
\usepackage{acro}
\usepackage[dvipsnames]{xcolor}
\DeclareAcronym{DNN}{
  short={DNN},
  long={Deep Neural Network}
}

\DeclareAcronym{HAD}{
  short={HAD},
  long={Holonic Active Distillation}
}

\DeclareAcronym{ML}{
  short={ML},
  long={Machine Learning}
}

\DeclareAcronym{WALT}{
  short={WALT},
  long={Watch and Learn Time-lapse}
}

\DeclareAcronym{TTA}{
  short={TTA},
  long={Test-Time Adaptation}
}

\DeclareAcronym{HoL}{
  short={HoL},
  long={Holonic Learning}
}

\DeclareAcronym{MAS}{
  short={MAS},
  long={Multi-Agent System}
}

\DeclareAcronym{HMAS}{
  short={HMAS},
  long={Holonic Multi-Agent System}
}

\DeclareAcronym{OMAS}{
  short={OMAS},
  long={Organizational Multi-Agent System}
}

\DeclareAcronym{CCTV}{
  short={CCTV},
  long={Closed-Circuit Television}
}

\DeclareAcronym{KD}{
  short={KD},
  long={Knowledge Distillation}
}

\DeclareAcronym{SBAL}{
  short={SBAL},
  long={Stream-Based Active Learning}
}

\DeclareAcronym{AL}{
  short={AL},
  long={Active Learning}
}

\DeclareAcronym{SBAD}{
  short={SBAD},
  long={Stream-Based Active Distillation}
}

\DeclareAcronym{CSBAD}{
  short={CSBAD},
  long={Clustered Stream-Based Active Distillation}
}

\DeclareAcronym{UDA}{
  short={UDA},
  long={Unsupervised Domain Adaptation}
}

\DeclareAcronym{MC}{
  short={MC},
  long={Model Compression}
}
\usepackage{amsmath}
\usepackage{xspace}
\usepackage{algorithm}
\usepackage{algpseudocode}
\usepackage{tabularx}
\usepackage{booktabs}
\newcommand{\name}[1]{\textbf{\textit{#1}}}

\newcommand{\eg}{\unskip \textit{e.g.},\xspace}
\newcommand{\ie}{\unskip \textit{i.e.},\xspace}
\newcommand{\N}{\ensuremath{N}\xspace}
\newcommand{\Np}{\ensuremath{N_+}\xspace}
\usepackage{orcidlink}
\usepackage{changes}

\usepackage[numbers]{natbib}

\begin{document}
\title{Holonic Active Distillation for Scalable Multi-Agent Learning in Multi-Sensor Systems}
%
%
\author{Dani Manjah\inst{1}\orcidlink{0000-0001-9034-0794} \and
Tim Bary\inst{1} \orcidlink{0009-0005-8333-8051}  \and
 Benoit Macq \inst{1}\orcidlink{0000-0002-7243-4778} \and
 Stéphane Galland\inst{2}\orcidlink{0000-0002-1559-7861}}

\authorrunning{D. Manjah et al.}

\titlerunning{HAD for Scalable Multi-Agent Learning}
%
\institute{Institute of Information and Communication Technologies, Electronics and Applied Mathematics (ICTEAM), UCLouvain, 1348 Louvain-la-Neuve, Belgium \\
\email{\{dani.manjah, tim.bary, benoit.macq\}@uclouvain.be}\and
Universit\'e de Technologie de Belfort Montb\'eliard, UTBM, CIAD UR 7533, F-90010 Belfort cedex, France\\
\email{stephane.galland@utbm.fr}\\}
\maketitle              
\begin{abstract}

The rapid expansion of sensor-based networks introduces major challenges in scalability, adaptability, and knowledge transfer, especially in open environments where new subsystems can dynamically join or leave. In this work, we propose a Holonic Active Distillation architecture within a \ac{HMAS} to address these issues. Our approach integrates \ac{CSBAD}, a framework in which specialized student models collect local data, query pseudo-labels from teacher models, and cluster into groups of similar sensors.

Results show that the holonic organization balances local specialization with global generalization, while efficiently adapting to sensor departures and re-integrations. We also analyzed trade-offs among incremental model updates, system reorganization, and scalability limits.

Our findings highlight the advantages of holonic learning for multi-sensor systems while identifying key challenges related to model drift and long-term adaptation.
\keywords{Holonic Multi-Agent Systems \and Distributed Learning \and Collaborative Learning \and Scalable Model Adaptation}
\end{abstract}

\section{Introduction}
In recent years, sensor systems have evolved from isolated and manageable units to expansive and interconnected networks \citep{Perera2014}.
This transformation, while enabling broader coverage, challenges the traditional approach of deploying a single, universal \ac{DNN} across all sensors \citep{CSBAD}.
The dynamic nature of real-world deployments, characterized by stochastic changes and the continuous addition of new sensors, introduces diverse contexts that demand ever larger training datasets and models \citep{yolov8}.
This upscaling not only incurs significant costs, but also accumulates hidden technical debt, complicating the maintenance required to adapt to distribution shifts in sensor data \citep{MLHiddenDebt,selfAdaptoDeviceChanges}.

 \begin{figure}[htpb]
     \centering
     \includegraphics[width=\linewidth]{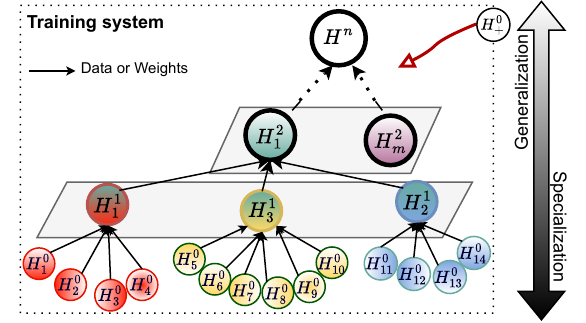}
     \caption{A large-scale distributed training system where \ac{DNN} nodes are trained on datasets built from a specificity-diversity trade-off for effective learning. Lower nodes are typically tailored to their task (\ie analytics on a sensor) and operationally more efficient. Higher nodes, trained over vaster and more diverse data, provide generalization ability. This system can scale up or down, causing challenges in integration and adaptability.}
     \label{fig:openLearningThreeLevel}
 \end{figure}
 
Traditional methods have relied on centralized, monolithic \acp{DNN} that struggle to scale with the increasing complexity and diversity of sensor networks.
These approaches often lead to inefficiencies and increased maintenance challenges.
Recent advances, such as the \ac{HoL} framework, offer a promising alternative by embracing the agent paradigm to improve scalability and flexibility \citep{esmaeili2023holonic}.
\ac{HoL} leverages self-nested structures of agents, known as holons, to integrate local and global perspectives, facilitating easier subsystem integration and preventing the propagation of disturbances
\citep{Koestler67,HolonicSelfIntegrationOfSubsystems,Rodriguez2011}.
This hierarchical learning approach improves the efficiency of data and algorithm handling, particularly for large distributed datasets \citep{HiearchicalLearning}.
A question remains related to the design of a scalable distributed learning framework that supports continuous \ac{DNN} refinement with minimal refactoring, while allowing each unit to update itself online, self-organize with its peers, and transfer knowledge as the system scales.

This work builds upon the \ac{HoL} framework by introducing strategies for aggregation, communication, and commitment between learning holons.
The contributions are twofold:
\begin{itemize}
\item We augment \ac{HoL} with standardized organizations and roles, allowing newcomers to integrate knowing only their role and the associated protocol, inspired by active learning and distillation of knowledge \citep{Ahmed_Abbas_2015,Rodriguez2011,CSBAD}.
\item We propose holonification mechanisms in which agents cluster horizontally and vertically based on the similarity of their sensor streams, preserving confidentiality and balancing specificity and diversity.
This recursive process improves the accuracy and resilience of the system, allowing seamless expansion or contraction without disruption \citep{diversityinholonic,largeScaleAnalysisMAS}.
\end{itemize}

This further advances the adaptability of multisensor holonic systems \citep{Manjah2024Autonomous},  materializing continuous learning in dynamic environments. 

The remainder of this paper is structured as follows. Section~ \ref{sec:sotaMultiAgentLearning} provides background and current related works. Section~\ref{sec:HADA} presents the high-level formulation of the organizations for the \ac{HAD} architecture and the relationships between super- and subholons. Section~\ref{sec:HoL} introduces a distributed, multi-tiered, self-nested structure for \acp{DNN} and the self-organization mechanisms that realize the specificity–diversity trade-off. Section~\ref{sec:materialsHolonic} describes the materials used in the experiment, while Section~\ref{sec:ResultsHOLearn} reports the results on sensor addition and removal, comparing partial reorganizations with complete retraining and measuring knowledge-transfer speed. Finally, Section~\ref{sec:Discussion} discusses the insights, limitations, and perspectives, followed by the conclusion in Section~\ref{sec:agentConclusions}. 
\section{Related Works}
We set the foundational machine learning frameworks to continuously adapt to incoming data. Next, we discuss the landscape of learning in~\ac{MAS} motivated by the ability of agent-oriented design to minimize refactoring and support isolated updates. The section is then concluded by reviewing organizational methodologies for seamless integration of a new sensor, model, or more generally, a subsystem of sensors and models in \ac{MAS}.

\subsection{Online Test-time Adaptation}

Adapting machine learning models at test-time is crucial to preserve their performance \cite{app10031154,Artus,ElliotBrion}. \ac{SBAD}~\citep{sbad} addresses the challenges of the \textbf{lack of labeled data at test time} by selecting, from sensor streams, samples to constitute a training set representative of the sensor's characteristics,\;\ie features space. To alleviate annotation costs, \ac{SBAD} relies on annotation by other models within the system in a \textbf{Teacher-Student} scheme. 
Fine-tuning a model per device does not scale well, as it requires maintaining a separate model for each additional sensor~\citep{fowler2018refactoring,MLHiddenDebt}. 

Clustered SBAD~\cite{CSBAD} diminishes the number of models by grouping sensors using a similarity distance of their features and by training one model per cluster. This also improves the accuracy of models by striking a balance between tailoring models to their data streams and generalization capabilities by training on enough diverse samples~\citep{gradientDiversity}. CSBAD's limitation is not to retain knowledge for successive re-training of current models, nor provide knowledge transfer mechanisms to new sensors.  

Inspired by human organizations, this paper proposes a design where higher layers of systems build structural knowledge \textbf{to seamlessly integrate agents and avoid the pitfalls of abrupt failures, environmental changes, or knowledge loss,} especially with new or varied data classes \citep{catastrophicforgettingfrench1999,FederatedContinuousLearning}.

\subsection{Learning in Multi-Agent Systems}
\label{sec:sotaMultiAgentLearning}
Adaptive \ac{MAS} networks leverage online learning strategies to dynamically respond to environmental changes, highlighting the importance of distributed and collaborative learning \citep{smartcities5010019}.

\citet{Wolpert1997} introduce a system utilizing reinforcement learning to align agent actions with collective goals, minimizing human oversight. Agents are organized into ``sub-worlds'' for focused collaboration, yet the application of reinforcement learning in complex scenarios with varied sensors and methods encounters obstacles such as unclear rewards and limited exploration, which hinders the required diversity of learning \citep{Wolpert1997,emergentSoftware,DiverseAutoCurriculum}.

Organizational learning considers agents evolving through both personal and collective learning efforts, enhancing agents' abilities in \ac{MAS} through management mediated interactions and task alignment to boost system efficiency \citep{socialLearningOrganizational,organizationallearningClassic}. This model emphasizes the role of knowledge sharing in improving workflows and establishing structural knowledge, crucial for system resilience. Social science research \citep{socialscienceopensystems} reflects on the applicability of this framework to understand the impact of staff turnover on management, analogous to agent dynamics in open MAS.

Hierarchical learning \citep{HierarchicalLearning} uses hierarchical MAS to streamline model training in various geographical locations. 
By modeling challenges as a hypergraph, the system organizes agents, each with unique skills and knowledge, into a structured multitiered network. 
This design not only facilitates the decentralized handling of \ac{ML} algorithms and data, but also significantly improves the efficiency and scalability of processing large distributed datasets.

In the context of distributed \ac{ML}, \citet{GUPTA20181} pioneered Federated Learning (FL) to train neural networks in distributed datasets, prioritizing data privacy and computational efficiency. However, FL faces hurdles in communication and training reliability. 
Hierarchical FL addresses these by grouping users to improve FL security and efficiency through group-specific updates \citep{HierarchicalFL}. 
Personalized FL \citep{PFL} methods aim to produce personalized models for different users or groups of users \citep{ClusteredPFL} to keep track of their individualized requirements. Hierarchy has also been instrumental in Fog Learning. 
Unlike FL, which is based on a star topology of device-server interactions, Fog Learning explicitly considers the network and topology structures among devices and enables intelligent device collaborations through data and parameter offloading \citep{FOL}. 

\citet{esmaeili2023holonic} abstract FL and CSBAD with \ac{HoL}, applying holonic principles to a collaborative learning framework. In that sense,\textbf{ FL and CSBAD can be seen as a first-order \ac{HoL}}. \ac{HoL} enhances model cooperation with specific strategies for aggregation, communication, and commitment within holons, facilitating complex yet intuitive collaboration of nodes compared to Fog Learning. In this balance between local autonomy and coordinated decision making, holonic systems are better equipped to tackle challenges such as adaptability, and scalability.

\ac{HoL} \textbf{does not specify how learning agents should (re)organize}, nor how a system can seamlessly expand to new domains or safely unlearn obsolete ones; shortcomings that become acute in applications requiring auto-scaling and auto-tuning \citep{esmaeili2023holonic,selfadaptsurvey}.

\subsection{Organizational Multi-Agent Systems}
Agent-oriented software engineering addresses the limitations of traditional methods like UML in managing complex, distributed, expanding and self-adaptive systems \citep{Ahmed_Abbas_2015,Wautelet2021-pb,Manjah2024Autonomous,HolonicSelfIntegrationOfSubsystems,selfAdaptoDeviceChanges,selfdevcollectivechallenges,selfadaptsurvey,Emas12Gleizes}. Organizational theory from social science inspired software designers who developed methodologies for the development of \ac{MAS} to break down design complexity via 1) the use of metaphors that are more accessible to software engineers and 2) a focus on high levels of abstraction to enable the integration of new agents, even when they differ significantly from existing ones~\cite{kolp2006multi,SCHATTEN2014576,garcia2008issues}. In fact, \acp{MAS} are best viewed as organizational structures of autonomous, proactive agents interacting to achieve shared or individual goals~\cite{kolp2006multi}. 

Many agent-oriented methodologies have been developed last decades, such as, ADELFE~\cite{bernon:hal-01205160,Emas12Gleizes}, ASPECS~\cite{Cossentino2010}, Gaia~\cite{Wooldridge2000}, INGENIAS~\cite{PavonIngenias2003}, PASSI~\cite{CossentinoPassi2005}, Pro-metheus~\cite{Prometheus2003}, SODA~\cite{OmiciniSoda2001}, Tropos~\cite{Tropos2002}, MOISE~\cite{MOISE}.
Each has its own specificities: ADELFE is dedicated to adaptive system and cooperative agents design, ASPECS is dedicated to holonic multi-agent systems, Gaia focuses on static organization and roles, whereas
PASSI focuses on agent social aspects thanks to ontology, SODA highlights the notion of environment. {MOISE focuses on explicit organisational modelling—defining roles, groups, missions, and deontic norms—to balance agent autonomy and coordinated behaviour.}

Given our choice of the holonic learning paradigm, we adopt the ASPECS methodology \citep{Cossentino2010}.

\subsection{Conclusion}
We address scalability limitations of continuous adaptation in machine learning systems by enriching the holonic learning paradigm with \emph{organizations} and \emph{roles}. These concepts provide an abstract interaction pattern that improves the architecture’s robustness and flexibility. Furthermore, our \emph{Teacher–Student} stream-based distillation scheme supplies pseudo-labels that calibrate online incoming \emph{Students}, thereby enabling auto-tuning. Finally, self-organization emerges from a specificity–diversity trade-off among \emph{Students}, while integration and deletion protocols dynamically scale sensor subsystems. Collectively, these mechanisms yield \textbf{the first \ac{HoL} variant that supports self-organization and auto-scaling.}

\section{Holonic Active Distillation Architecture}
\label{sec:HADA}

We seek a design that minimizes refactoring and supports isolated updates, simplifying the integration of a new sensor, model, or more generally a subsystem of sensors and models \citep{fowler2018refactoring,MLHiddenDebt}. From the literature, we derived five main recommendations to \textbf{design scalable, multimethod learning systems}:

\begin{enumerate}
    \item Establish standardized interaction protocols, aggregation strategies, commitment, and communication patterns within components. This facilitates the integration of new units, as they only need to understand their role and communication methods within the system, regardless of their operating mode \citep{Ahmed_Abbas_2015,Rodriguez2011}.
    \item Render a method, a sensor, or by construction a subsystem as independent and self-contained as possible to limit the complexity between units. This aims to simplify a local update or maintenance \citep{Ahmed_Abbas_2015,Emas12Gleizes,Wautelet2021-pb}. 

    \item Recursively divide a system into subsystems based on a key criterion.  This prevents the propagation of disturbances \citep{Rodriguez2011}. Furthermore, integration of a new component requires less communication as it requires only coordination with the upper layers of the system instead of with each subsystem \citep{HolonicSelfIntegrationOfSubsystems}.
    \item Place units at certain levels of the hierarchy and provide representations of how other levels can contribute ``information'' or ``models''. This division simplifies the complexity of programming, allowing designers to focus on each module and facilitating reuse between different systems \citep{HolonicSelfIntegrationOfSubsystems}.

    \item Exploit \emph{Active Distillation}, where each \emph{Student} unit collects data on the fly from its streams to train on them. Training is performed by querying a model \emph{Teacher} \citep{sbad}.

\end{enumerate}

Given our choice of the organizational holonic paradigm, we adopt the ASPECS methodology \citep{Cossentino2010}. The latter starts by defining an \name{Organization}, which denotes a subsystem in which components play a role and interact to achieve a shared goal in the context of this organization. Next, the \name{Roles} which are both expected behaviors to fulfill (part of) requirements, and status to the role's agent in the organization (Section~\ref{sec:learningOrganization}). The subsequent activity (Section~\ref{sec:holarchy}) is the definition of relationships between superholons (higher-level entities) and subholons (lower-level entities). As a reference later, a \name{holarchy} denotes the hierarchy of self-regulating holons.
\subsection{Teacher-Specialized Student}
\label{sec:learningOrganization}

\begin{figure}[htpb]
  \centering
    \includegraphics[width=0.75\linewidth]{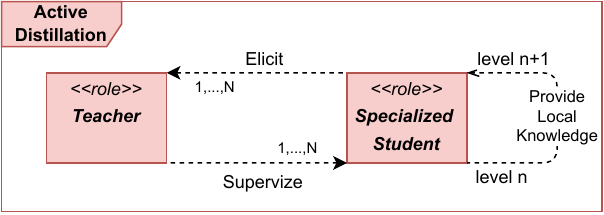}
    \caption{Organizational model of the \name{Teacher-Student}, using the ASPECS notation \cite{Cossentino2010}. The \name{Specialized Student} role involves a component tasked with building expertise over a delineated sub-domain in the system, \; \ie a regional distribution. The \name{Teacher} role supervizes the learning processes of the \emph{Students}.}
    \label{fig:LearningOrganisation}
\end{figure}

Building on the Active Distillation framework and the specificity-diversity trade-off\index{Specificity-diversity trade-off} from \citep{CSBAD}, we developed an organizational model that incorporates the roles of \name{Specialized Student} and \name{Teacher}, as illustrated in Fig.~\ref{fig:LearningOrganisation}.

The \name{Specialized Student} role is designed to continuously collect data on subparts of the system's deployment environment. Under the oversight of a higher-order \name{Teacher} entity, these \emph{Students} learn from these data, adapting their models' weights accordingly.

\subsection{Holarchy}
\label{sec:holarchy}

The section begins with introducing a new notation. Then we present an example of a three-tiered holarchy structure. Each level of this holarchy is a possible instance of an organization defined in Section~\ref{sec:learningOrganization}. To provide a more holistic perspective, we depict the Cyber-Physical Platform (CPP) data processing organization (see our previous work \citep{Manjah2024Autonomous}) alongside the TSS but CPP is not the main focus of this paper\footnote{As a more detailed context, \name{CPP} is designed to respond to external requests with perceptions and to manage its finite resources to ensure fair access across multiple surveillance operations. The \name{Resource Provider} role ensures a fair distribution of the resources among all parties. The \name{Observer} role has the ability to produce perceptions thanks to the data acquired by the \name{Sensor} role. The data acquisition could be based on another CPP.}.

\subsubsection{Notations}\label{sec:holonNotations}
A holarchy \( \mathcal{H}^L_{\name{O}} \) includes up to \( L \) vertical layers instantiating an organization \name{O}. A holon \( i \) in layer \( l \), where \( l \) ranges from \( 0 \) to \(L \), is denoted by \( \hbar^l_i \)  and comprises:
\begin{itemize}
    \item \( \mathbf{\mathrm{X}}^l_i \): Set of operating data streams of a holon \( \hbar^l_i \).
    \item \( \mathcal{T}^l_i \): Training set of a holon \( \hbar^l_i \).
    \item \( \mathcal{V}^l_i \): Validation set of a holon \( \hbar^l_i \).
    \item \( \theta^{\hbar^l_i} \): Processing model of a holon \( \hbar^l_i \).
    \item \( \mathsf{SUB}^l_i \): Inner members corresponding to layer \( l-1 \) of a holon \( \hbar^l_i \).
    \item \( \mathsf{SUP}^l_i \): Superior holon of a holon \( \hbar^l_i \).
\end{itemize}

\subsubsection{Multi-Scale Hierarchical Architecture}
\label{sec:HMASdesign}

\begin{figure}[htpb]
\centering
\includegraphics[width=\linewidth]{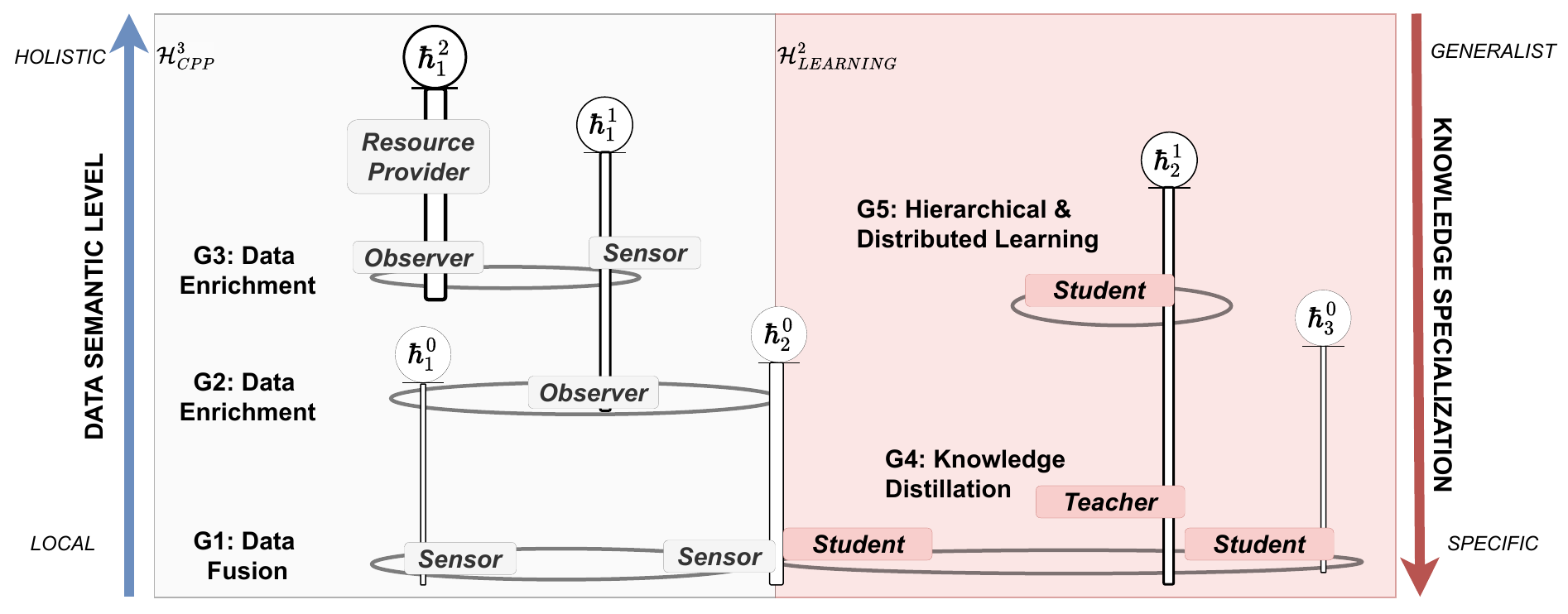}
\caption{Holonic architecture inspired by the ``cheese board'' notation \citep{Cossentino2010,FERAUD2017}. Each level represents a different hierarchical position, defining both the semantic level of data and the degree of knowledge specialization. On the left, the $\mathcal{H}^3_{\name{CPP}}$ instantiates the \name{CPP} organization, and on the right, the $\mathcal{H}^2_{\name{TS}}$ is responsible for active learning. Agents may assume multiple roles and participate in multiple holarchies simultaneously.}
\label{fig:cheeseBoard}
\end{figure}

The system architecture, shown in Fig.~\ref{fig:cheeseBoard}, includes two holarchies: $\mathcal{H}^3_{\name{CPP}}$ that processes data on three levels and $\mathcal{H}^2_{\name{TSS}}$ managing knowledge on two levels.

\begin{itemize}
    \item\textbf{At level 0:} agents are the primary functional layer. They employ models designed for specific data streams. Proximity to other agents, geographically or related to the task, allows them to merge outputs and reduce errors. For example, $\hbar^0_1$ and $\hbar^0_2$ form Group G1 to fuse their outputs to feed the data request of a higher-order holon $\hbar^1_1$.

    However, agents monitoring the same area may employ different models if their functions require learning different features. Consequently, $\hbar^0_2$ and $\hbar^0_3$, as \name{Specialized Students}, form Group G4 to learn a shared model under the supervision of a \name{Teacher} holon via Group G5.
    
    \item\textbf{At levels 1 and 2:} higher levels above the agents integrate and synthesize data from specific areas of the system (\eg data streams that share attributes). The holons in the role \name{Observer}, such as $\hbar^1_1$ and $\hbar^2_1$, elevate the collected data to a new semantic level. 
    
    Holon $\hbar^1_2$, a higher order \name{Specialized Student}, aggregates validation sets from $\hbar^0_2$ and $\hbar^0_3$ (\ie \(\mathcal{T}^1_2 = \mathcal{V}^0_2 \cup \mathcal{V}^0_3\)), to create a broader and generalized model.
\end{itemize}

Generally, each semantic level consolidates knowledge across \textbf{broader areas of the system, fostering a holistic view, such as city-scale tracking.} Meanwhile, intermediate layers consisting of \name{Specialized Students} synthesize knowledge from lower levels, to \textbf{deepen collective task understanding} and \textbf{increase holons' universality}.   
\section{Holonic Learning Framework}
\label{sec:HoL}

This section presents mechanisms to create a multilevel learning framework. Next, it introduces a mechanism to incorporate new nodes by coordinating with the top layers and assigning each new node to the group whose \ac{DNN} model is most accurate in its data stream. 

\subsection{Holonification}
\label{sec:holonification}
In the holonic terminology, \name{holonification} is the process of grouping agents into a holarchy, resembling complex clustering based on criteria like capabilities and resource access \citep{diversityinholonic,largeScaleAnalysisMAS}.

In this Section, we propose a multi-tiered learning structure (illustrated in Fig.~\ref{fig:openLearningThreeLevel}), comprising a portfolio of models that range from sensor-specific to universal, deployable across the entire network. Specifically, upper-layer models are trained on larger datasets for broader coverage, while lower-layer models use smaller, more similar datasets for increased specificity. Having intermediate models at various levels of granularity not only ensures adaptability, \textbf{but also supports robust knowledge organization.} For example, city-wide vehicle detection may require multiple models specializing in certain domain representations \citep{CSBAD}. However, these domain-specific models \textbf{benefit from interactions with peer models or a more fundamental model} that develops a fundamental understanding of object detection tasks \citep{gradientDiversity}. Our agent-based modeling offers this flexibility to develop these vertical and horizontal interactions. 

Formally, each holon in a layer $l > 0$ is allocated a budget \( B^l\ = B_0 \cdot 10^l\), where $B_0$ represents the number of images used for model fine-tuning. This budget limits the training of each layer to at most $10^l$ from the preceding levels, ensuring that the size of the data set of any holon \(\hbar^l_i\) does not exceed \( B^l\),\;\ie \(|\mathcal{T}^l_i|\leq B^l\).

To merge holons, we adopt the premise from \citet{CSBAD} that models with similar performance have learned from comparable data.  

The remainder of this section describes the holonification process.

\paragraph{STEP 1 \textendash \; Cross-Performance Vector.}
Assuming holons can transfer their model weights to each other within the same layer. Each holon \( \hbar^l_i \) computes a performance vector \(P_i\) by evaluating the effectiveness of models from other holons and itself in the same layer on its own validation data \( \mathcal{V}^{l}_i \), according to Equation~\ref{eq:crossperformancevector}.
\begin{equation}
\label{eq:crossperformancevector}
    P^{l}_i := \left[\begin{array}{c}
    
    f\left(\theta^{l}_1, \mathcal{V}^{l}_{i}\right); 
    \cdots ;
    f\left(\theta^{l}_{N^l}, \mathcal{V}^{l}_{i}\right)
    \end{array}\right]^{T} 
\end{equation}

\noindent Where \(\theta_j\) denotes the model parameters of the \(j\)-th holon $\hbar^l_j$, $j \in 1,\cdots, N^l$ and \( f(\theta, \mathcal{V}) \)  the score of a model performance \( \theta \) against a validation set \( \mathcal{V} \). 

\paragraph{STEP 2 \textendash \; Pair-Wise Distance Computation.}

To quantify the differences between models trained in different domains, holons broadcast their cross-perfor-mance vectors \( P^l_i \) defined in Equation~\ref{eq:crossperformancevector} and compute a pairwise distance between their performance in the datasets and the performance of the other holons. Generally, for a holon \(\hbar^l_i\), the distance to a holon \(\hbar^l_j\) is given by Equation \ref{eq:distancematrixHolonic}.
\begin{equation}
\label{eq:distancematrixHolonic}
D_i(\hbar^l_j) = \sqrt{\sum_{k=1}^{N^l} (P^l_{ik} - P^l_{jk})^2}
\end{equation}

\paragraph{STEP 3 \textendash \; Agglomerative Merging using Single Linkage.}
The merging of the holons is an iterative process. The set of holons \(\{\hbar^l_1,\cdots,\hbar^l_{N^l} \}\) creates a higher-order holarchy $\mathcal{H}^{l+1}$ to which they belong. 

At each iteration, the set of holons \(\hbar^{l}\) transmits their smallest linkage distance. This is defined as the minimal distance between the inner members of the holons. Formally, for two holons \( \hbar^{l}_A \) and \( \hbar^{l}_B \), the single link distance \( L(\hbar^{l}_A, \hbar^{l}_B) \) is given by Equation \ref{eq:linkageHolonic}.
\begin{equation}
\label{eq:linkageHolonic}
L(\hbar^{l}_A, \hbar^{l}_B) = \min \{ D_{ij} : \hbar^{l-1}_i \in \mathsf{SUB}(\hbar^{l}_A), \; \hbar^{l-1}_j \in \mathsf{SUB}(\hbar^{l}_B) \}
\end{equation}

After all linkages are evaluated, the pair with the smallest distance merges, involving a combination of their datasets. After merging, the set of holons has decreased, \(\{\hbar^l_1,\cdots,\hbar^l_{N^l-1} \}\), and the linkage distances are updated for all agents. 

The process ends if there remains only one holon or if the previous merge leads to a holon with a dataset size that exceeds \(B^{l+1}\). In the second scenario, the process goes back to STEP 1 for the set \(\{\hbar^{l+1}_1,\cdots,\hbar^l_{N^{l+1}}\}\).

\paragraph{STEP 4 \textendash \; Model Training.}
The final steps consist in training the cluster models on the aggregated data sets.

\subsection{Domain Integration Process}\label{sec:integration}

A new holon $\hbar_+$ joins a holarchy $\mathcal{H}^L$ of $L$ levels. Its integration starts at the highest hierarchical level, $L$, and progresses downward to the level $1$. At each level, $\hbar_+$ is associated with the holon $\hbar^1_*$ that shows the highest performance in the new set of unit validations, $\mathcal{V}_+$, subject to meeting budget constraints, ~\ie $ |\mathcal{T}^{l}_* \cup \mathcal{T}_+| \leq B^l$. Once integrated, $\hbar_+$'s dataset merges with that of the selected holon, $\hbar^l_*$, necessitating a retraining of the aggregated dataset. If no appropriate holons are available at a required level, the system can initiate reholonification, integrating $\hbar_+$ with the set \( \{\hbar^l_1,\cdots,\hbar^l_{N^l}\}\). A pseudocode is provided in Algorithm~\ref{alg:on-the-fly integration}.

\begin{algorithm}[htpb]
    \caption{Integration of a New Holon into a Holarchy}
    \label{alg:on-the-fly integration}
    \begin{algorithmic}[1] 
    \Require $\mathcal{H}^L$: $L$-level learning holarchy
    \Require $\hbar_+$: A holon
    
    \For{$l = L$ downto $1$}
        \Statex \Comment{\textbf{Identify sub-holons whose training sets do not exceed budget constraints}}
        
        \State $\mathsf{FreeHolons} \gets { \hbar^{l}  : |\mathcal{T}^{l} \cup \mathcal{T}_+| \leq B^l }$
        \If{$\mathsf{FreeHolons} = \emptyset $}

        \State Reholonification with the set \( \{\hbar^l_1,\cdots,\hbar^l_{N^l}\} \cup \hbar_+\).
        
        \State \textbf{break}
        \EndIf
        
        \Statex \Comment{\textbf{Select the optimal sub-holon for integration}}
        \State $\mathsf{FreeModels} \gets  \mathsf{FreeHolons}\text{'s models}$
        
        \State $\theta^{l}_{*} \gets \arg \max_{\;\theta^{l} \in \mathsf{FreeModels}} f\left(\theta^{l}, \mathcal{V}_+\right)$
        
        \State Update model parameters $\theta^{l}_*$ using $\mathcal{T}^{l}_* \cup \mathcal{T}_+$.
        
        \Statex \Comment{\textbf{Integrate $\hbar_+$ into holon of \(\theta^l_*\)}}
        
         \State $\mathsf{SUB}(\hbar^l_*) \gets \mathsf{SUB}(\hbar^l_*) \cup {\hbar_+}$
    \EndFor
    \end{algorithmic}
\end{algorithm}

\begin{remark}
\textbf{The cost of holonification} is compared to integration on-the-fly on the basis of the amount of communication between the holons. It is built on a single linkage-Hierarchical Clustering, with a \textbf{complexity of $\mathcal{O}\left(\N^2\right)$} \citep{SLINK}. 

On the other hand, the \textbf{on-the-fly mechanism has a $\mathcal{O}(\N+L)$ complexity.} This corresponds to the worst-case scenario in which the new agent is compared to all holons from the upper layer \( L \) to layer 0. This mechanism thus offers a cost-effective integration in comparison with a reholonification. 
\end{remark}

\subsection{Research Questions}
From the setup and challenges described above, we formulate the following research questions:

\begin{enumerate}
    \item Given a new data stream, how can we determine the most suitable existing model for fine-tuning?
    \item Assume an effective integration of new sensors based on similarity with a group of sensors from the system: \begin{enumerate}
        \item What are the consequences on model accuracy upon the integration of a new agent?
        \item How does the accuracy of the model scale when incrementally integrating \(\Np\) new agents versus performing a full system reorganization?
    \end{enumerate}
    \item What are the long-term accuracy trade-offs between retaining versus discarding data from removed sensors?
\end{enumerate}
\section{Materials and Methods}
\label{sec:materialsHolonic}

The datasets, the training procedure, and the evaluation protocol are presented in this Section. 

\subsection{Datasets}
We used two city-focused video datasets for a total of 16 cameras.

\paragraph{WALT~\citep{Reddy_2022_CVPR}}
features footage from nine static cameras over 1--4 weeks. Sampling rates vary (5,000--40,000 frames/week), with temporal bursts and diverse weather conditions (snow, rain, day/night).

\paragraph{AI-City~\citep{Naphade21AIC21}}
features seven annotated videos, each approximately five minutes at 10~FPS. Camera angles and sensor types vary (vertical, dome, PTZ), ensuring coverage of multiple representation contexts.

\subsection{Model Training}
We follow \ac{SBAD} \citep{CSBAD} sampling 256 images per camera. A large YOLOv8x6 Teacher (261.1 GFLOPs) pseudolabels these samples. 
Each Student model is a YOLOv8n (8.7 GFLOPs), initialized with COCO weights \citep{COCO}, then fine-tuned at a learning rate of 0.01 (unless otherwise indicated).

\subsection{Evaluation of mAP50-95}
We report the \emph{``mAP50-95"} as the \textbf{m}ean \textbf{A}verage \textbf{P}recision across various intersection over union thresholds, spanning from \textbf{0.50 to 0.95} in increments of 0.05 . We evaluated the holon's performance on its associated datasets.

\section{Results}
\label{sec:ResultsHOLearn}

We begin by evaluating our holonification approach under different budgets and then proceed with incremental integration, departure handling, and knowledge-transfer experiments.

\subsection{Holonification Baseline Performance}
\label{sec:holonificationResults}
We conducted a holonification\index{Holonification}, as proposed in Section~\ref{sec:holonification}, on a dataset comprising sixteen cameras. We set a multilayer budget framework \(B^l = 256 \cdot 10^l\), implying that layer 0 holons do not exceed 256 training samples, and successive layers cannot exceed $10^l$ sub-holons for a holon $\hbar^l$.  
\begin{table}[htpb]
    \caption{Holonification with varying budgets. Shown are the final groupings and average mAP50-95 for the 16-camera dataset.}
    \label{tab:HolonificationResults}
    \centering
    \footnotesize
    \begin{tabularx}{\linewidth}{lccX}
    \toprule
    \textbf{Layer} &\(\mathbf{B^l}\) & \textbf{mAP50-95} & \textbf{Holonic Structure} \\
    \midrule
    2&25600 & 0.65& \(\mathcal{H}^2:\{\hbar^1_0, \hbar^{1}_{1},\hbar^{1}_{2}\}\) \\
    \addlinespace
    \midrule
    
    1 &2560 & 0.66 & \(\begin{aligned} 
                                     &\mathcal{H}^1_0:\{\hbar^0_0, \hbar^{0}_{1}, \hbar^{0}_{2},\hbar^{0}_{8}\} \\
                                     \addlinespace
                                     &\mathcal{H}^1_1:\{\hbar^0_3, \hbar^{0}_{4}, \hbar^{0}_{5},\hbar^{0}_{6},\hbar^{0}_{7}\} \\
                                     \addlinespace
                                     &\mathcal{H}^1_2:\{\hbar^0_{9}, \hbar^{0}_{10}, \hbar^{0}_{11},\hbar^{0}_{12},\hbar^{0}_{13},\hbar^{0}_{14},\hbar^{0}_{15}\}
                         \end{aligned}\) \\
    \midrule
    0&256 & 0.67 & \(\mathcal{H}^1:\{\hbar^0_0,\ldots, \hbar^{0}_{15}\}\) \\
    \bottomrule
    \addlinespace
    \(\text{YOLOv8n}^{\text{COCO}}\)& N.A&  0.498 & N.A\\
    \addlinespace
    \bottomrule
    \end{tabularx}
\end{table}
Table~\ref{tab:HolonificationResults} confirms that the models require specificity to achieve maximum performance.

\subsection{Transfer, Integration and Departure}
\label{sec:OpenLearn}
\subsubsection{Model Transfer Upon Increment}
The transferability of holons across new, although similar, domains is investigated. Table \ref{tab:psanalysisanticlustering} details the performance results for models trained in an all-but-one combination of domains as well as across all domains. 

\begin{table}[htpb]
    \centering
    \caption{mAP50-95 scores for models trained under an all-but-one camera to assess the transferability of those models on the remaining camera. A baseline is also provided where the model is trained across all cameras. Each model is trained for \(10 000\) iterations.}
    \label{tab:psanalysisanticlustering}
    \begin{tabularx}{\linewidth}{X*{4}{r}}
        \toprule
        & \multicolumn{4}{c}{\(f(\theta, \mathcal{V}_i)\)} \\
        \cmidrule(lr){2-5}
        \textbf{Cluster} & \(\mathcal{V}^0_0\) & \(\mathcal{V}^0_1\) & \(\mathcal{V}^0_2\) & \(\mathcal{V}^0_3\) \\
        \midrule
        
        $\theta^{\mathcal{H}^1_1 \; \backslash \; \{\hbar^0_0\}}$ 
        & \textcolor{Maroon}{\textit{0.38}} & 0.65 & 0.65 & 0.47 \\

        $\theta^{\mathcal{H}^1_1 \; \backslash \; \{\hbar^0_1\}}$
        & 0.46 & \textcolor{Maroon}{\textit{0.57}} & \textcolor{ForestGreen}{\textbf{0.66}} & 0.46 \\

        $\theta^{\mathcal{H}^1_1 \; \backslash \; \{\hbar^0_2\}}$
        & \textcolor{ForestGreen}{\textbf{0.49}} & \textcolor{ForestGreen}{\textbf{0.66}} & \textcolor{Maroon}{\textit{0.65}} & \textcolor{ForestGreen}{\textbf{0.48}} \\

        $\theta^{\mathcal{H}^1_1 \; \backslash \; \{\hbar^0_8\}}$
        & 0.47 & 0.64 & 0.66 & \textcolor{Maroon}{\textit{0.42}} \\
        \midrule
        $\theta^{\mathcal{H}^1_1}$  
        & 0.46 & 0.65 & 0.66 & 0.47 \\
        \bottomrule
    \end{tabularx}
\end{table}

The results indicate that models struggle to transfer, even across similar camera domains, reinforcing the need to integrate the newcomer in a cluster, and the local retraining the cluster. 

\subsubsection{Incremental Integration}
\label{sec:incrementalIntegrationResults}
We evaluated the impact of integrating \Np new units into a holonified system, structured with budget limits of
\(B^l= 256 \cdot 10^l\). Using our integration mechanism described in Algorithm~\ref{alg:on-the-fly integration},
we evaluated two scenarios: integrating one (\Np = 1) and three (\Np = 3) additional agents.

In 16 agent configurations, the incremental integration maintained an average mAP50-95 of $0.66 \pm 0.003$ (\Np = 1) and $0.66 \pm 0.006$ (\Np = 3), showing no degradation compared to the baseline in Table~\ref{tab:HolonificationResults}.

\subsubsection{Agent Departure}
\begin{figure}[htpb]
    \centering
    \includegraphics[width=0.80\linewidth]{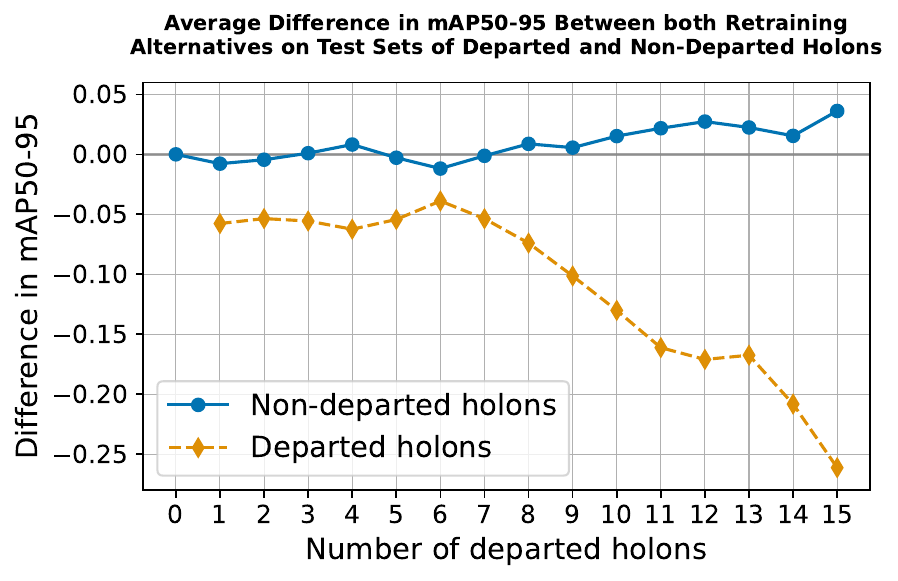}
    \caption{Difference in \(\hbar^2\) model performance between retaining and discarding each departed sensor's data. Blue: remaining sensors; yellow: departed sensors. 
    Results show marginal gains for remaining sensors but a marked degradation on departed sensors.}
\label{fig:gracefully_deg}
\end{figure}

When a sensor $\hbar^0_i$ leaves, its data $\mathcal{T}_i$ may be retained or discarded. 
We successively simulate the exit of each agent and track the accuracy of the global model on (i) remaining and (ii) left sensors.
As Fig.~\ref{fig:gracefully_deg} shows, removing a sensor's data yields small gains for the remaining sensors, but severely reduces performance if that sensor later re-enters the system. Note that, upon the departure of an agent, their data set \(\mathcal{T}^0_i\) is removed from the collective data set \(\mathcal{T}^2\), and the model is re-trained for 10,000 iterations at a learning rate of 0.005.

\subsection{Inter-Holonic Knowledge Transfer}
\label{sec:InterHolonic}
\begin{figure}[htbp]
    \centering
    \includegraphics[width=0.8\linewidth]{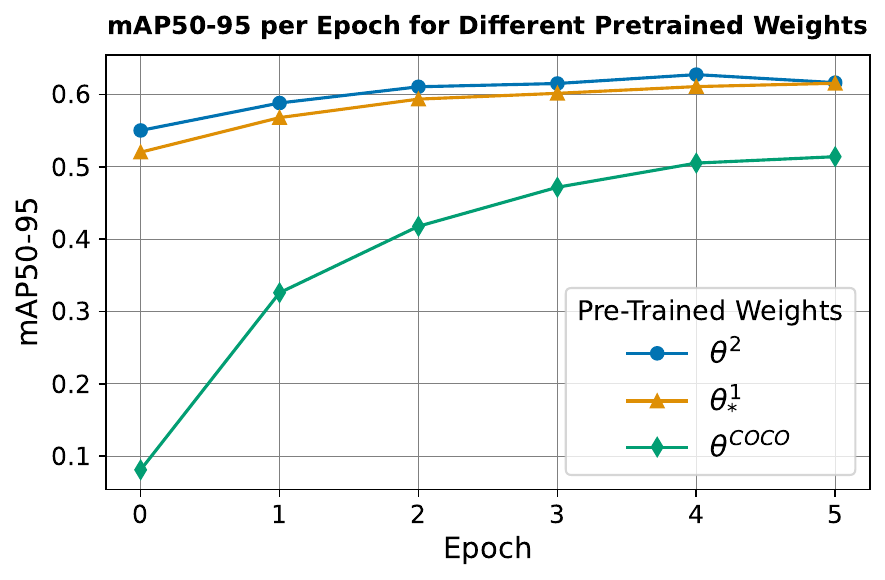}
    \caption{mAP50-95 per epoch for a new model starting from universal model \(\theta^{2}\), group-specific \(\theta^{1}_*\), and general-purpose \(\theta^{\text{COCO}}\). The superiority of \(\theta^{2}\) highlights the efficiency of selecting a pretrained model closer to the source.}
    \label{fig:training_perf}
\end{figure}

We test how effectively a holon trained on existing cameras can accelerate training and improve the peak accuracy of a newcomer domain. Specifically, we conducted 16 trials, each excluding one camera from the data set to simulate a ``newcomer''. The following pre-trained models serve as initial weights:
\begin{itemize}
    \item $\theta^2$: Global holon (trained on 15 cameras),
    \item $\theta^1_*$: Group-specific holon,
    \item YOLOv8n$^\text{COCO}$: General-purpose off-the-shelf model.
\end{itemize}
Fig.~\ref{fig:training_perf} shows that $\theta^2$ or $\theta^1_*$ consistently outperform the generic COCO baseline when fine-tuning the newcomer camera. Training spanned 5 epochs with a learning rate of 0.005.

\section{Discussion}
\label{sec:Discussion}
\subsection{Insights}
Our experiments confirm that a certain level of domain specificity improves accuracy (Section~\ref{sec:holonificationResults}), though it also increases the number of models to maintain. Budget constraints help contain this growth but can reduce performance gains from specialized holons. Meanwhile, leveraging broader universal models accelerates learning for new domains (Section~\ref{sec:InterHolonic}). 

In the context of open systems (Section~\ref{sec:OpenLearn}), our sanity check shows that a straightforward model transfer performs under, even when the model comes from similar domains.
The observed performance gap motivated the development of integration mechanisms, which proved effective, but the experiments do not provide conclusive evidence regarding the maximum number of agents that can be integrated without a performance decline.
Finally, agent departure highlights a trade-off between short-term gains and relearning costs if the environment reappears. In other words, discarding data should be considered in terms of agents' turnover rate.

\subsection{Limitations}
\paragraph{Machine Learning Lifecycle}
Machine learning based models are also subject to feedback loops, where data and interactions with the external world influence their behavior in unintentional ways \citep{MLHiddenDebt}. The subsequent design of machine learning systems should account for the fact that their behavior evolves with environmental data and user interactions. This includes providing control mechanisms to avoid the accumulation of errors due to the self-supervised nature of the system. 

\paragraph{Stress-tests} We need to further stress test the system; that is, starting with a system of size $N$, stress tests can evaluate how many new components ($\Np$) can be integrated without compromising the quality of service.

\subsection{Perspectives}

Modern systems integrate heterogeneous approaches (\eg physics-based modeling vs. deep learning) and diverse sensing modalities (\eg cameras, radar), producing richer analytics \citep{CAMPAGNER2023241,Manjah2024Autonomous}. Our architecture abstracts the holonic paradigm sufficiently to accommodate such heterogeneity. However, specialized coordination modules could further optimize collaborative performance among different modalities.

\section{Conclusions}
\label{sec:agentConclusions}
We present an organizational holonic learning design coupled with active learning to address the challenges of scaling learning in multisensor networks. 
Our self-organization mechanism, grounded in the specificity-diversity trade-off, allows for the establishment of various granularity levels and handles sensor addition and removal while maintaining strong predictive performance. 
Experimental results highlight the benefits of vertical and horizontal knowledge transfer, although more stress testing is needed to refine the upper limits on system growth. We also note that self-supervised processes risk model drift without robust monitoring, which may cause issues in autoscaling and autotuning. Future work aims to design colearning mechanisms for heterogeneous methods. 

\begin{credits}
\subsubsection{\ackname} This work was partially funded by PIT ATMP - Convention 8881. T. Bary is funded by the MedReSyst project, supported by FEDER and the Walloon Region.

\subsubsection{\discintname}
The authors have no competing interests to declare that
are relevant to the content of this article.
\end{credits}
\bibliographystyle{splncs04nat}
\bibliography{EMAS20205_13_references}

@InProceedings{Emas12Gleizes,
author="Gleizes, Marie-Pierre",
title="Self-adaptive Complex Systems",
booktitle="Multi-Agent Systems",
year="2012",
publisher="Springer",
address="Berlin, Heidelberg",
pages="114--128",
isbn="978-3-642-34799-3"
}

@InProceedings{bernon:hal-01205160,
author="Bernon, Carole
and Gleizes, Marie-Pierre
and Peyruqueou, Sylvain
and Picard, Gauthier",
editor="Petta, Paolo
and Tolksdorf, Robert
and Zambonelli, Franco",
title="ADELFE: A Methodology for Adaptive Multi-agent Systems Engineering",
booktitle="Engineering Societies in the Agents World III",
year="2003",
publisher="Springer Berlin Heidelberg",
address="Berlin, Heidelberg",
pages="156--169",
DOI = {10.1007/3-540-39173-8_12},
isbn="978-3-540-39173-9"}

@article{FERAUD2017,
title = "First Comparison of SARL to Other Agent-Programming Languages and Frameworks",
journal = "Procedia Computer Science",
volume = "109",
pages = "1080 - 1085",
year = "2017",
issn = "1877-0509",
doi = "https://doi.org/10.1016/j.procs.2017.05.389",
author = "Maxime Feraud and Stéphane Galland"
}

@article{kolp2006multi,
  title={Multi-agent architectures as organizational structures},
  author={Kolp, Manuel and Giorgini, Paolo and Mylopoulos, John},
  journal={Autonomous Agents and Multi-Agent Systems},
  volume={13},
  pages={3--25},
  year={2006},
  publisher={Springer}
}

@article{SCHATTEN2014576,
title = {Towards a Formal Conceptualization of Organizational Design Techniques for Large Scale Multi Agent Systems},
journal = {Procedia Technology},
volume = {15},
pages = {576-585},
year = {2014},
note = {2nd International Conference on System-Integrated Intelligence: Challenges for Product and Production Engineering},
issn = {2212-0173},
doi = {10.1016/j.protcy.2014.09.018},
author = {Markus Schatten and Petra Grd and Mladen Konecki and Robert Kudelić},
}

@inproceedings{garcia2008issues,
  title={Issues for organizational multiagent systems development},
  author={Garcia, Emilia and Argente, Estefania and Giret, Adriana and Botti, Vicente},
  booktitle={Sixth International Workshop From Agent Theory to Agent Implementation (AT2AI-6)},
  pages={59--65},
  year={2008},
  organization={Citeseer}
}

@inproceedings{MLHiddenDebt,
 author = {Sculley, D. and Holt, Gary and Golovin, Daniel and Davydov, Eugene and Phillips, Todd and Ebner, Dietmar and Chaudhary, Vinay and Young, Michael and Crespo, Jean-Fran\c{c}ois and Dennison, Dan},
 booktitle = {Advances in Neural Information Processing Systems},
 pages = {},
 publisher = {Curran Associates, Inc.},
 title = {Hidden Technical Debt in Machine Learning Systems},
 volume = {28},
 year = {2015}
}

@book{fowler2018refactoring,
  title={Refactoring: improving the design of existing code},
  author={Fowler, Martin},
  year={2018},
  publisher={Addison-Wesley Professional}
}

@article{Wolpert1997,
author = {Wolpert, David H. and Macready, William G.},
doi = {10.1109/4235.585893},
issn = {1089778X},
journal = {IEEE Transactions on Evolutionary Computation},
number = {1},
pages = {67--82},
title = {{No free lunch theorems for optimization}},
volume = {1},
year = {1997}
}

@article{Rodriguez2011,
author = {Rodriguez, Sebastian and Hilaire, Vincent and Gaud, Nicolas and Galland, Stephane and Koukam, Abderrafi{\^{a}}a},
doi = {10.1007/978-3-642-17348-6_11},
journal = {Natural Computing Series},
pages = {251--279},
publisher = {Springer, Berlin, Heidelberg},
title = {{Holonic Multi-Agent Systems}},
volume = {37},
year = {2011}
}

@article{Ahmed_Abbas_2015,
	doi = {10.11648/j.ijiis.20150403.11},
	year = 2015,
	publisher = {Science Publishing Group},
	volume = {4},
	number = {3},
	pages = {46},
	author = {Hosny Ahmed Abbas},
	title = {Organization of Multi-Agent Systems: An Overview},
	journal = {International Journal of Intelligent Information Systems},
}

@INCOLLECTION{Wautelet2021-pb,
  title     = "Agent-based software engineering, paradigm shift, or research
               program evolution",
  booktitle = "Research Anthology on Recent Trends, Tools, and Implications of
               Computer Programming",
  author    = "Wautelet, Yves and Schinckus, Christophe and Kolp, Manuel",
  publisher = "IGI Global",
  pages     = "1642--1654",
  year      =  2021
}

@article{CAMPAGNER2023241,
title = {Aggregation models in ensemble learning: A large-scale comparison},
journal = {Information Fusion},
volume = {90},
pages = {241-252},
year = {2023},
issn = {1566-2535},
doi = {https://doi.org/10.1016/j.inffus.2022.09.015},
author = {Andrea Campagner and Davide Ciucci and Federico Cabitza}
}

@InProceedings{Naphade21AIC21,
author = {Milind Naphade and Shuo Wang and David C. Anastasiu and Zheng Tang and Ming-Ching Chang and Xiaodong Yang and Yue Yao and Liang Zheng and Pranamesh Chakraborty and Christian E. Lopez and Anuj Sharma and Qi Feng and Vitaly Ablavsky and Stan Sclaroff},
title = {The 5th AI City Challenge},
booktitle = {The IEEE Conference on Computer Vision and Pattern Recognition (CVPR) Workshops},
year = {2021},
}

@InProceedings{sbad,
    author    = {Manjah, Dani and Cacciarelli, Davide and Standaert, Baptiste and Benkedadra, Mohamed and de Hertaing, Gauthier Rotsart and Macq, Beno{\^\i}t and Galland, St\'ephane and De Vleeschouwer, Christophe},
    title     = {Stream-Based Active Distillation for Scalable Model Deployment},
    booktitle = {Proceedings of the IEEE/CVF Conference on Computer Vision and Pattern Recognition (CVPR) Workshops},
    year      = {2023},
    pages     = {4998-5006}
}

@article{SBAL,
   author = {Davide Cacciarelli and Murat Kulahci and John Sølve Tyssedal},
   doi = {10.1016/j.knosys.2022.109664},
   issn = {09507051},
   journal = {Knowledge-Based Systems},
   pages = {109664},
   title = {Stream-based active learning with linear models},
   volume = {254},
   year = {2022},
}

@InProceedings{Reddy_2022_CVPR,
    author    = {Reddy, N. Dinesh and Tamburo, Robert and Narasimhan, Srinivasa G.},
    title     = {WALT: Watch and Learn 2D Amodal Representation From Time-Lapse Imagery},
    booktitle = {Proceedings of the IEEE/CVF Conference on Computer Vision and Pattern Recognition (CVPR)},
    year      = {2022},
    pages     = {9356-9366}
}

@article{COCO,
  author    = {Tsung{-}Yi Lin and
               Michael Maire and
               Serge J. Belongie and
               Lubomir D. Bourdev and
               Ross B. Girshick and
               James Hays and
               Pietro Perona and
               Deva Ramanan and
               Piotr Doll{\'{a}}r and
               C. Lawrence Zitnick},
  title     = {Microsoft {COCO:} Common Objects in Context},
  journal   = {CoRR},
  volume    = {abs/1405.0312},
  year      = {2014},
}

@misc{yolov8,
author = {Jocher, Glenn and Chaurasia, Ayush and Qiu, Jing},
license = {AGPL-3.0},
title = {{Ultralytics YOLO}},
url = {https://github.com/ultralytics/ultralytics},
version = {8.0.0},
year = {2023}
}

@article{catastrophicforgettingfrench1999,
title = {Catastrophic forgetting in connectionist networks},
journal = {Trends in Cognitive Sciences},
volume = {3},
number = {4},
pages = {128-135},
year = {1999},
issn = {1364-6613},
doi = {10.1016/S1364-6613(99)01294-2},
author = {Robert M. French}
}

@article{SLINK,
    author = {Sibson, R.},
    title = "{SLINK: An optimally efficient algorithm for the single-link cluster method}",
    journal = {The Computer Journal},
    volume = {16},
    number = {1},
    pages = {30-34},
    year = {1973},
    issn = {0010-4620},
}

@book{Koestler67,
    title = {The ghost in the machine},
    author = {Koestler, Arthur},
    publisher = {Hutchinson},
    address = {London, UK},
    year = 1967,
}

@article{Wooldridge2000,
author = {Wooldridge, Michael and Jennings, Nicholas R. and Kinny, David},
doi = {10.1023/A:1010071910869},
journal = {Autonomous Agents and Multi-Agent Systems},
pages = {285--312},
publisher = {Springer},
title = {The Gaia Methodology for Agent-Oriented Analysis and Design},
volume = {3},
number = {3},
year = {2000}
}

@InProceedings{PavonIngenias2003,
author="Pav{\'o}n, Juan
and G{\'o}mez-Sanz, Jorge",
title="Agent Oriented Software Engineering with INGENIAS",
booktitle="Multi-Agent Systems and Applications III",
year="2003",
publisher="Springer",
address="Berlin, Heidelberg",
pages="394--403",
isbn="978-3-540-45023-8"
}

@InProceedings{CossentinoPassi2005,
author="Cossentino, Massimo
and Gaglio, Salvatore
and Sabatucci, Luca
and Seidita, Valeria",
title="The PASSI and Agile PASSI MAS Meta-models Compared with a Unifying Proposal",
booktitle="Multi-Agent Systems and Applications IV",
year="2005",
publisher="Springer Berlin Heidelberg",
address="Berlin, Heidelberg",
pages="183--192",
isbn="978-3-540-31731-9"
}

@InProceedings{Prometheus2003,
author="Padgham, Lin
and Winikoff, Michael",
title="Prometheus: A Methodology for Developing Intelligent Agents",
booktitle="Agent-Oriented Software Engineering III",
year="2003",
publisher="Springer Berlin Heidelberg",
address="Berlin, Heidelberg",
pages="174--185",
isbn="978-3-540-36540-2"
}

@InProceedings{OmiciniSoda2001,
author="Omicini, Andrea",
title="SODA: Societies and Infrastructures in the Analysis and Design of Agent-Based Systems",
booktitle="Agent-Oriented Software Engineering",
year="2001",
publisher="Springer Berlin Heidelberg",
pages="185--193",
isbn="978-3-540-44564-7"
}

@article{Tropos2002,
title = {Towards requirements-driven information systems engineering: the Tropos project},
journal = {Information Systems},
volume = {27},
number = {6},
pages = {365-389},
year = {2002},
issn = {0306-4379},
doi = {10.1016/S0306-4379(02)00012-1},
author = {Jaelson Castro and Manuel Kolp and John Mylopoulos}
}

@article{Cossentino2010,
author={Cossentino, Massimo
and Gaud, Nicolas
and Hilaire, Vincent
and Galland, St{\'e}phane
and Koukam, Abderrafi{\^a}a},
title={ASPECS: an agent-oriented software process for engineering complex systems},
journal={Autonomous Agents and Multi-Agent Systems},
year={2010},
volume={20},
number={2},
pages={260-304},
doi={10.1007/s10458-009-9099-4}}

@INPROCEEDINGS{emergentSoftware,
  author={Porter, Barry and Rodrigues Filho, Roberto},
  booktitle={2019 IEEE 13th International Conference on Self-Adaptive and Self-Organizing Systems (SASO)}, 
  title={Distributed Emergent Software: Assembling, Perceiving and Learning Systems at Scale}, 
  year={2019},
  volume={},
  number={},
  pages={127-136}}

@inproceedings{DiverseAutoCurriculum,
author = {Yang, Yaodong and Luo, Jun and Wen, Ying and Slumbers, Oliver and Graves, Daniel and Bou Ammar, Haitham and Wang, Jun and Taylor, Matthew E.},
title = {Diverse Auto-Curriculum is Critical for Successful Real-World Multiagent Learning Systems},
year = {2021},
isbn = {9781450383073},
publisher = {International Foundation for Autonomous Agents and Multiagent Systems},
address = {Richland, SC},
booktitle = {Proceedings of the 20th International Conference on Autonomous Agents and MultiAgent Systems},
pages = {51–56},
numpages = {6},
location = {Virtual Event, United Kingdom},
series = {AAMAS '21}
}

@article{selfadaptsurvey,
author = {Weyns, Danny and Gerostathopoulos, Ilias and Abbas, Nadeem and Andersson, Jesper and Biffl, Stefan and Brada, Premek and Bures, Tomas and Di Salle, Amleto and Galster, Matthias and Lago, Patricia and Lewis, Grace and Litoiu, Marin and Musil, Angelika and Musil, Juergen and Patros, Panos and Pelliccione, Patrizio},
title = {Self-Adaptation in Industry: A Survey},
year = {2023},
issue_date = {June 2023},
publisher = {Association for Computing Machinery},
address = {New York, NY, USA},
volume = {18},
number = {2},
issn = {1556-4665},
journal = {ACM Trans. Auton. Adapt. Syst.},
articleno = {5},
numpages = {44}
}

@INPROCEEDINGS{selfAdaptoDeviceChanges,
  author={Beal, Jacob and Viroli, Mirko and Pianini, Danilo and Damiani, Ferruccio},
  booktitle={2016 IEEE 10th International Conference on Self-Adaptive and Self-Organizing Systems (SASO)}, 
  title={Self-Adaptation to Device Distribution Changes}, 
  year={2016},
  volume={},
  number={},
  pages={60-69},
  doi={10.1109/SASO.2016.12}}

@INPROCEEDINGS{HolonicSelfIntegrationOfSubsystems,
  author={Diaconescu, Ada and Frey, Sylvain and Müller-Schloer, Christian and Pitt, Jeremy and Tomforde, Sven},
  booktitle={2016 IEEE 10th International Conference on Self-Adaptive and Self-Organizing Systems (SASO)}, 
  title={Goal-Oriented Holonics for Complex System (Self-)Integration: Concepts and Case Studies}, 
  year={2016},
  volume={},
  number={},
  pages={100-109},
  doi={10.1109/SASO.2016.16}}

@article{socialscienceopensystems,
title = {Should knowledge be distorted? Managers' knowledge distortion strategies and organizational learning in different environments},
journal = {The Leadership Quarterly},
volume = {32},
number = {3},
pages = {101477},
year = {2021},
issn = {1048-9843},

author = {Jiamin Dong and Renjing Liu and Yu Qiu and Mary Crossan}
}

@article{Perera2014,
author = {Perera, Charith and Zaslavsky, Arkady and Christen, Peter and Georgakopoulos, Dimitrios},
issn = {2161-3915},
journal = {Transactions on Emerging Telecommunications Technologies},
number = {1},
pages = {81--93},
publisher = {John Wiley {\&} Sons, Ltd},
title = {{Sensing as a service model for smart cities supported by Internet of Things}},
volume = {25},
year = {2014}
}

@article{HierarchicalLearning,
author = {Esmaeili, Ahmad and Gallagher, John C. and Springer, John A. and Matson, Eric T.},
title = {HAMLET: A Hierarchical Agent-based Machine Learning Platform},
year = {2022},
issue_date = {December 2021},
publisher = {Association for Computing Machinery},
address = {New York, NY, USA},
volume = {16},
number = {3–4},
issn = {1556-4665},
doi = {10.1145/3530191},
journal = {ACM Trans. Auton. Adapt. Syst.},
month = jul,
articleno = {9},
numpages = {46}
}

@inproceedings{esmaeili2023holonic,
author = {Esmaeili, Ahmad and Ghorrati, Zahra and Matson, Eric T.},
title = {Holonic Learning: A Flexible Agent-based Distributed Machine Learning Framework},
year = {2024},
isbn = {9798400704864},
publisher = {International Foundation for Autonomous Agents and Multiagent Systems},
address = {Richland, SC},
booktitle = {Proceedings of the 23rd International Conference on Autonomous Agents and Multiagent Systems},
pages = {525–533},
numpages = {9},
location = {Auckland, New Zealand},
series = {AAMAS '24}
}

@article{GUPTA20181,
title = {Distributed learning of deep neural network over multiple agents},
journal = {Journal of Network and Computer Applications},
volume = {116},
pages = {1-8},
year = {2018},
issn = {1084-8045},
doi = {10.1016/j.jnca.2018.05.003},
author = {Otkrist Gupta and Ramesh Raskar}
}

@inproceedings{ClusteredPFL,
 author = {Ghosh, Avishek and Chung, Jichan and Yin, Dong and Ramchandran, Kannan},
 booktitle = {Advances in Neural Information Processing Systems},
 pages = {19586--19597},
 publisher = {Curran Associates, Inc.},
 title = {An Efficient Framework for Clustered Federated Learning},
 volume = {33},
 year = {2020}
}

@ARTICLE{FOL,
  author={Hosseinalipour, Seyyedali and Brinton, Christopher G. and Aggarwal, Vaneet and Dai, Huaiyu and Chiang, Mung},
  journal={IEEE Communications Magazine}, 
  title={From Federated to Fog Learning: Distributed Machine Learning over Heterogeneous Wireless Networks}, 
  year={2020},
  volume={58},
  number={12},
  pages={41-47}}

@ARTICLE{PFL,
  author={Ma, Zhenguo and Xu, Yang and Xu, Hongli and Liu, Jianchun and Xue, Yinxing},
  journal={IEEE Transactions on Mobile Computing}, 
  title={Like Attracts Like: Personalized Federated Learning in Decentralized Edge Computing}, 
  year={2024},
  volume={23},
  number={2},
  pages={1080-1096},
  ISSN={1558-0660}}

@InProceedings{largeScaleAnalysisMAS,
author="Esmaeili, Ahmad
and Mozayani, Nasser
and Jahed-Motlagh, Mohammad Reza
and Matson, Eric T.",
title="Towards Topological Analysis of Networked Holonic Multi-agent Systems",
booktitle="Advances in Practical Applications of Survivable Agents and Multi-Agent Systems: The PAAMS Collection",
year="2019",
publisher="Springer International Publishing",
address="Cham",
pages="42--54",
isbn="978-3-030-24209-1"
}

@article{diversityinholonic,
title = {The impact of diversity on performance of holonic multi-agent systems},
journal = {Engineering Applications of Artificial Intelligence},
volume = {55},
pages = {186-201},
year = {2016},
issn = {0952-1976},
author = {Ahmad Esmaeili and Nasser Mozayani and Mohammad Reza Jahed Motlagh and Eric T. Matson}
}

@inproceedings{Manjah2024Autonomous,
  author    = {D. Manjah and S. Galland and C. De Vleeschouwer and B. Macq},
  title     = {Autonomous Methods in Multisensor Architecture for Smart Surveillance},
  booktitle = {Proceedings of the 16th International Conference on Agents and Artificial Intelligence},
  volume    = {3},
  pages     = {824--832},
  year      = {2024},
  isbn      = {978-989-758-680-4},
  issn      = {2184-433X},
}

@InProceedings{gradientDiversity,
  title = 	 {Gradient Diversity: a Key Ingredient for Scalable Distributed Learning},
  author = 	 {Yin, Dong and Pananjady, Ashwin and Lam, Max and Papailiopoulos, Dimitris and Ramchandran, Kannan and Bartlett, Peter},
  booktitle = 	 {Proceedings of the Twenty-First International Conference on Artificial Intelligence and Statistics},
  pages = 	 {1998--2007},
  year = 	 {2018},
  volume = 	 {84},
  series = 	 {Proceedings of Machine Learning Research},
  publisher =    {PMLR},
}

@ARTICLE{FederatedContinuousLearning,
  author={Le, Junqing and Lei, Xinyu and Mu, Nankun and Zhang, Hengrun and Zeng, Kai and Liao, Xiaofeng},
  journal={IEEE Transactions on Cybernetics}, 
  title={Federated Continuous Learning With Broad Network Architecture}, 
  year={2021},
  volume={51},
  number={8},
  pages={3874-3888},
  doi={10.1109/TCYB.2021.3090260},
  ISSN={2168-2275},
  }

@article{MOISE,
  title={Instrumenting multi-agent organisations with organisational artifacts and agents: “Giving the organisational power back to the agents”},
  author={H{\"u}bner, Jomi F and Boissier, Olivier and Kitio, Rosine and Ricci, Alessandro},
  journal={Autonomous agents and multi-agent systems},
  volume={20},
  number={3},
  pages={369--400},
  year={2010},
  publisher={Springer}
}

@incollection{socialLearningOrganizational,
title = {Learning: Organizational},
booktitle = {International Encyclopedia of the Social \& Behavioral Sciences (Second Edition)},
publisher = {Elsevier},
edition = {Second Edition},
address = {Oxford},
pages = {695-698},
year = {2015},
isbn = {978-0-08-097087-5},
author = {Silvia Gherardi}
}

@InProceedings{organizationallearningClassic,
author="Terabe, Masahiro
and Washio, Takashi
and Katai, Osamu
and Sawaragi, Tetsuo",
title="A study of organizational learning in multiagents systems",
booktitle="Distributed Artificial Intelligence Meets Machine Learning Learning in Multi-Agent Environments",
year="1997",
publisher="Springer Berlin Heidelberg",
address="Berlin, Heidelberg",
pages="168--179",
isbn="978-3-540-69050-4"
}

@Article{smartcities5010019,
AUTHOR = {Nezamoddini, Nasim and Gholami, Amirhosein},
TITLE = {A Survey of Adaptive Multi-Agent Networks and Their Applications in Smart Cities},
JOURNAL = {Smart Cities},
VOLUME = {5},
YEAR = {2022},
NUMBER = {1},
PAGES = {318--347},
ISSN = {2624-6511}
}

@article{HierarchicalFL,
  title={Adaptive hierarchical federated learning over wireless networks},
  author={Xu, Bo and Xia, Wenchao and Wen, Wanli and Liu, Pei and Zhao, Haitao and Zhu, Hongbo},
  journal={IEEE Transactions on Vehicular Technology},
  volume={71},
  number={2},
  pages={2070--2083},
  year={2021},
  publisher={IEEE}
}

@INPROCEEDINGS{selfdevcollectivechallenges,
  author={Lippi, Marco and Mariani, Stefano and Martinelli, Matteo and Zambonelli, Franco},
  booktitle={2022 17th Conference on Computer Science and Intelligence Systems (FedCSIS)}, 
  title={Individual and Collective Self-Development: Concepts and Challenges}, 
  year={2022},
  volume={},
  number={},
  pages={15-21}}

@article{CSBAD,
title = {Camera clustering for scalable stream-based active distillation},
journal = {Expert Systems with Applications},
volume = {290},
pages = {128408},
year = {2025},
issn = {0957-4174},
doi = {https://doi.org/10.1016/j.eswa.2025.128408},
author = {Dani Manjah and Davide Cacciarelli and Christophe {De Vleeschouwer} and Benoît Macq}
}

@inproceedings{ElliotBrion,
author = {Eliott Brion and Jean L{\'e}ger and Umair Javaid and John Lee and Christophe De Vleeschouwer and Benoit Macq},
title = {{Using planning CTs to enhance CNN-based bladder segmentation on cone beam CT}},
volume = {10951},
booktitle = {Medical Imaging 2019: Image-Guided Procedures, Robotic Interventions, and Modeling},
editor = {Baowei Fei and Cristian A. Linte},
organization = {International Society for Optics and Photonics},
publisher = {SPIE},
pages = {109511M},
year = {2019},
doi = {10.1117/12.2512791},
URL = {https://doi.org/10.1117/12.2512791}
}

@InProceedings{Artus,
author = {Cioppa, Anthony and Deliege, Adrien and Istasse, Maxime and De Vleeschouwer, Christophe and Van Droogenbroeck, Marc},
title = {ARTHuS: Adaptive Real-Time Human Segmentation in Sports Through Online Distillation},
booktitle = {Proceedings of the IEEE/CVF Conference on Computer Vision and Pattern Recognition (CVPR) Workshops},
month = {June},
year = {2019}}

@article{app10031154,
AUTHOR = {Léger, Jean and Brion, Eliott and Desbordes, Paul and De Vleeschouwer, Christophe and Lee, John A. and Macq, Benoit},
TITLE = {Cross-Domain Data Augmentation for Deep-Learning-Based Male Pelvic Organ Segmentation in Cone Beam CT},
JOURNAL = {Applied Sciences},
VOLUME = {10},
YEAR = {2020},
NUMBER = {3},
ARTICLE-NUMBER = {1154},
URL = {https://www.mdpi.com/2076-3417/10/3/1154},
ISSN = {2076-3417},
DOI = {10.3390/app10031154}
}

@article{HiearchicalLearning,
author = {Esmaeili, Ahmad and Gallagher, John C. and Springer, John A. and Matson, Eric T.},
title = {HAMLET: A Hierarchical Agent-based Machine Learning Platform},
year = {2022},
issue_date = {December 2021},
publisher = {Association for Computing Machinery},
address = {New York, NY, USA},
volume = {16},
number = {3–4},
issn = {1556-4665},
journal = {ACM Trans. Auton. Adapt. Syst.},
month = {jul},
articleno = {9},
numpages = {46}
}

\end{document}